# Sliding-Ferroelectric Switching of Valley Polarization and Layer-Locked Anomalous Valley Hall Effect

Xue-Bing Peng,[1,*] San-Dong Guo,[2,†] Chaobiao Zhou,[1] Mingsu Si,[3] and Daqiang Gao,[4]

[1]*School of Physics and Mechatronic Engineering, Guizhou Minzu University, Guiyang, 550025, China*

[2]*School of Electronic Engineering, Xi'an University of Posts and Telecommunications, Xi'an 710121, China*

[3]*School of Materials and Energy, Lanzhou University, Lanzhou 730000, China*

[4]*Key Laboratory of Magnetism and Magnetic Functional Materials (Lanzhou University), Ministry of Education, Lanzhou 730000, P. R. China*

[*]Contact author: pengxb1228@outlook.com

[†]Contact author: sandongyuwang@163.com

Valley polarization (VP) and anomalous valley Hall effect (AVHE) offer complementary functionalities for information encoding and electrical readout. However, achieving their joint switching without reversing the magnetic order or reconstructing the Berry-curvature distribution is highly desirable for low-power valleytronics but remains challenging in magnetic materials. Here, we propose a symmetry-guided sliding-ferroelectric paradigm in fully compensated A-type antiferromagnetic bilayers. Interlayer sliding lowers the point-group symmetry from $D_{3h}$ to $C_{3v}$ by breaking the horizontal mirror symmetry while preserving the threefold rotation, thereby allowing switchable out-of-plane polarization while forbidding in-plane polarization. The resulting layer potential couples to the valley-contrasting layer polarization, reversing their valley order without altering the antiferromagnetic order. Furthermore, the layer-valley locking with opposite Berry curvatures enables ferroelectric switching of layer-locked AVHE through valley-selective hole occupation, without modifying the Berry-curvature distribution. The combination of first-principles calculations and a Wannier-based tight-binding model on $VSe_2$ verifies these mechanisms and reveals a strong ferroelectric polarization of 0.68 pC/m and a giant VP of 71 meV. The ferroelectric switching of VP is robust across the considered correlation-strength range. These results provide a route toward nonvolatile valley encoding and reversible Hall readout in antiferromagnetic valleytronics.

## I. INTRODUCTION

The valley degree of freedom, associated with inequivalent energy extrema in momentum space, provides an additional avenue for information encoding and processing beyond charge and spin [1-4]. In two-dimensional (2D) materials, broken inversion symmetry can give rise to valley-contrasting Berry curvature (BC), while spin-orbit coupling (SOC) further enriches valley-dependent electronic properties [5-7]. Particularly, opposite BCs at inequivalent valleys generate transverse carrier velocities in response to an in-plane electric field, giving rise to the valley Hall effect [8-12]. In systems with time-reversal symmetry, the two time-reversal-related valleys have degenerate energies and opposite BCs, leading to cancellation of their contributions to the net charge Hall conductivity, although a finite valley Hall current can remain. When time-reversal symmetry is broken, this cancellation is no longer symmetry-enforced. If magnetic order lifts the valley degeneracy and the occupation-weighted BC contributions from the two valleys become uncompensated, a finite anomalous Hall conductivity (AHC) can emerge, giving rise to the anomalous valley Hall effect (AVHE) [13-19]. Thus, valley polarization (VP) and AVHE offer complementary functionalities for information encoding and electrical readout, making their realization and switching an important objective for valleytronic applications.

Various approaches have been developed to manipulate valley states, including circularly polarized optical excitation [20, 21], external magnetic fields [22, 23], magnetic proximity effects [24, 25], and electrostatic gating [18, 26]. However, each of these approaches faces different limitations. Circularly polarized light enables valley-selective carrier excitation through optical selection rules, but the resulting nonequilibrium valley population generally relaxes once illumination is removed. Magnetic fields can indirectly manipulate valley states through magnetic-valley locking, yet the induced VP is typically weak. Magnetic proximity effects require complicated interfacial magnetic structures that are difficult to operate. Electrostatic gating can tune valley energies through layer-dependent electrostatic potentials, but the induced effect is volatile. In this context, ferrovalley materials have attracted considerable attention because the intrinsic coupling between magnetism and valley degrees of freedom enables spontaneous VP and valley-dependent transverse transport [27-31]. Nevertheless, switching VP in this manner requires magnetization reversal, which simultaneously reconstructs BC and hence the associated Hall response. This not only entails higher switching energy and slower dynamics but also runs counter to the low-power, disturbance-resistant operation paradigm pursued by valleytronics [32, 33]. Therefore, developing a mechanism that enables joint switching of VP and AVHE without reversing the magnetic order or reconstructing the BC remains an urgent problem to be addressed.

Antiferromagnetic (AFM) materials offer an attractive alternative to conventional ferrovalley systems owing to their fully compensated magnetic moments, reduced stray fields, and potential for ultrafast magnetic dynamics [34, 35]. Particularly, A-type AFM bilayers [as schematically shown in Fig. S1(a) of Supplemental Material (SM)], with intralayer parallel and interlayer antiparallel magnetic alignment, provide an additional layer degree of freedom that can couple to spin and valley states, enabling control of valley-dependent transport while preserving the underlying magnetic configuration [36, 37]. Sliding ferroelectricity, distinct from conventional ferroelectricity driven primarily by ionic displacements, originates from relative in-plane translation between adjacent van der Waals (vdW) layers and can generate switchable polarization [38, 39]. The combination of sliding ferroelectricity and A-type AFM bilayers thus opens a new route to addressing the above problem. Recent studies have demonstrated coupling between ferroelectric polarization and valley degrees of freedom, suggesting the possibility of electrically controlling VP [40, 41]. However, the symmetry conditions governing sliding-induced polarization and the microscopic mechanism converting real-space ferroelectric polarization into momentum-space VP remain to be fully clarified. A layer-locked AVHE has also been proposed in an A-type AFM insulator via electrostatic gating, where the layer-dependent potential tunes the relative spin contributions to the Hall response [42]; however, this response remains volatile, and valley-relevant BC needs to be reversed. These advances raise an important question: can sliding ferroelectricity provide a nonvolatile route to simultaneously switch VP and the layer-locked AVHE while preserving the AFM order and retaining the BC distribution? Resolving this question requires understanding how symmetry-allowed ferroelectric polarization couples to layer-resolved valley states and subsequently controls transverse transport.

In this work, we investigate sliding-ferroelectric control of VP and the layer-locked AVHE in fully compensated A-type AFM bilayers of the $VX_2$ (X = S, Se, and Te) family using first-principles calculations and symmetry analysis. We demonstrate that interlayer sliding breaks the horizontal mirror symmetry while preserving the threefold rotational symmetry, reducing the structural point group from $D_{3h}$ to $C_{3v}$ at the polar stackings. This symmetry reduction allows switchable out-of-plane ferroelectric polarization while forbidding in-plane polarization. The resulting layer potential couples to the valley-contrasting layer polarization of the $K_1$ and $K_2$ states, shifting their energies in opposite directions and reversing their relative valley order without changing the AFM order. In $VSe_2$, this mechanism produces a substantial VP of approximately 71 meV, accompanied by a switchable polarization of approximately 0.68 pC/m. The opposite signs of polarization and VP in the two switching states remain robust across the investigated range of effective Hubbard interactions. Furthermore, calculations for $VS_2$ and $VTe_2$ bilayers reveal substantial VP of approximately 54 and 95 meV, respectively, demonstrating the applicability of this mechanism across the $VX_2$ family. Beyond valley switching, the reversal of valley order changes the occupation-weighted contributions from the two valleys with opposite BCs, thereby reversing the sign and layer character of AVHE while leaving the overall BC distribution largely preserved. These findings establish sliding ferroelectricity as a promising route to nonvolatile valley encoding and reversible, layer-selective Hall readout in compensated AFM materials.

## II. COMPUTATIONAL METHODS

First-principles calculations were performed using the Vienna ab initio Simulation Package (VASP) within density functional theory (DFT) [43, 44]. The exchange-correlation interaction between electrons was described by the generalized gradient approximation (GGA) with the Perdew–Burke–Ernzerhof (PBE) functional [45]. The projector augmented-wave (PAW) method was employed to describe the electron-ion interaction. Van der Waals interactions between top and bottom layers were treated using the DFT-D3 correction. A plane-wave kinetic-energy cutoff of 520 eV and a 15 × 15 × 1 Monkhorst–Pack $k$-point mesh were set for structural relaxation and self-consistent electronic-structure calculations. A vacuum region of 20 Å was introduced along the out-of-plane direction to eliminate interactions between periodic images. The atomic positions were relaxed until the residual forces on each atom were below 0.001 eV/Å, and the electronic self-consistency criterion was set to $10^{-7}$ eV. The strong correlation effects of the V-3$d$ orbitals were treated using the DFT+U approach with an effective on-site Hubbard parameter $U_{\text{eff}}$ = 1.0 eV [46]. Band structures were calculated along high-symmetry paths, with 200 points sampled between adjacent high-symmetry points. The ferroelectric polarization was calculated using the Berry-phase method [47, 48]. VASPKIT was used for post-processing the VASP calculations [49]. For the transport calculations, maximally localized Wannier functions were constructed using Wannier90 [50, 51], with V-$d$ and Se-$p$ orbitals included in the projection basis. The disentanglement procedure was employed, with an inner energy window from −5 to 3.5 eV and an outer energy window from −18 to 4 eV. A 200 × 200 × 1 $k$-point mesh was employed for BC and AHC calculations. The chemical-potential dependence of AHC was evaluated within the rigid-band approximation.

## III. RESULTS AND DISCUSSION

### A. Central mechanism

The central mechanism involves the coupling between sliding ferroelectricity and VP, which further enables a switchable layer-locked AVHE. As illustrated in Fig. 1(a), the two opposite polarized states, $+P_z$ and $-P_z$, are associated with opposite VP states. This association originates from the layer potential induced by interlayer sliding. In the A-type AFM bilayers, the valence-band-edge states at the inequivalent $K_1$ and $K_2$ valleys are predominantly localized in different layers, establishing pronounced layer-valley locking. The resulting layer potential, through its coupling to the valley-contrasting layer polarization, shifts the two valleys oppositely, thereby lifting their degeneracy and setting their relative energy order. Reversing the ferroelectric polarization therefore reverses the valley order and switches VP, while preserving both the underlying AFM order and the layer localization of each valley. This mechanism demonstrates that VP can be electrically switched through interlayer sliding without requiring magnetic-order reversal.

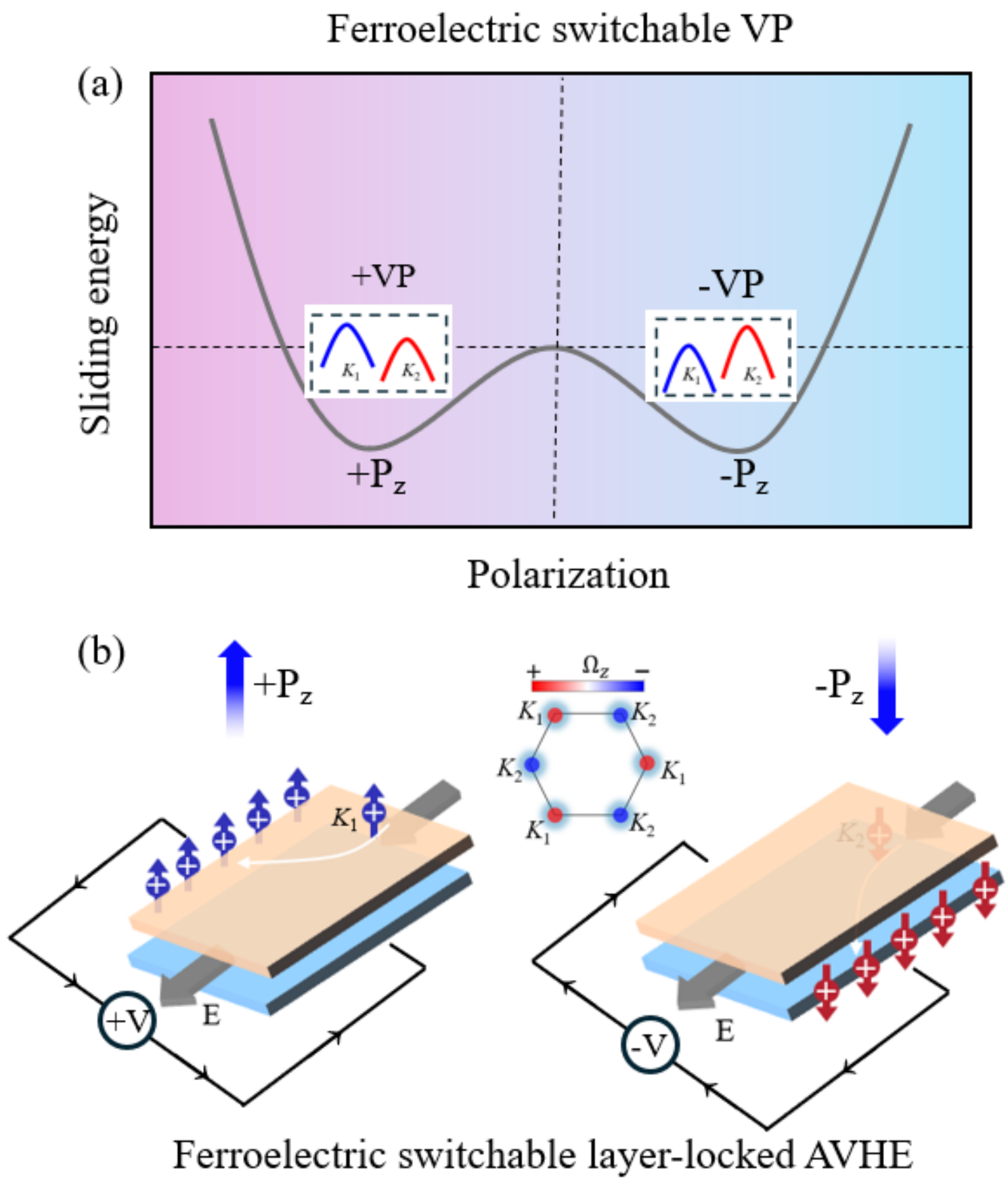

**FIG. 1.** Schematic illustration of ferroelectric switching of valley polarization (VP) and the layer-locked anomalous valley Hall effect (AVHE). (a) Two sliding-ferroelectric states with opposite polarizations, ($+P_z$) and ($-P_z$), exhibit reversed valence-band valley energy orderings, resulting in opposite VPs. (b) An in-plane electric field drives anomalous transverse carrier velocities through the valley-contrasting Berry curvature (BC). Because $K_1$ and $K_2$ states are predominantly localized in different layers, the valley-selective Hall response is layer locked. Reversing the ferroelectric polarization switches the dominant valley and reverses the Hall response, while the BC distributions at these two polarized states remain largely unchanged.

The resulting valley switching further enables a reversible layer-locked AVHE, as schematically illustrated in Fig. 1(b). Under an in-plane electric field, the opposite BCs at $K_1$ and $K_2$ generate transverse carrier velocities with opposite signs. Because these valleys are predominantly localized in different layers, their transverse transport contributions are also layer-resolved. When the chemical potential enters the VP window, the unequal valley occupation leads to incomplete cancellation of their occupation-weighted BC contributions, producing a finite anomalous Hall conductivity. Reversing the ferroelectric polarization switches the relative valley order and consequently changes the valley contribution to the Hall response. Since the two valleys carry opposite BCs and are associated with different layers, this switching reverses both the sign of the anomalous Hall conductivity and its layer contribution. Importantly, the Hall reversal occurs without reconstruction of the Berry-curvature distribution; it is primarily driven by the change in valley-selective occupation. Thus, sliding ferroelectricity provides a unified route to switching VP and the associated layer-locked AVHE while preserving the underlying AFM order and the essential BC landscape.

## B. Symmetry analysis of sliding ferroelectricity

To investigate the proposed switching mechanism, we construct bilayers composed of two identical $VX_2$ (X = S, Se, and Te) monolayers stacked along the perpendicular direction. Since isostructural compounds sharing the same space group but differing in the chalcogen element generally exhibit similar properties, we summarize their structural parameters and key electronic properties in Table I and focus on the experimentally accessible $VSe_2$ as a representative case for the following discussion. Full structural relaxation of bilayer $VSe_2$ gives $a = b = 3.34$ Å, $\gamma = 120°$, and an interlayer distance $d = 3.24$ Å, characterizing a van der Waals material with a regular hexagonal lattice. The binding energy, defined as $E_b = E_{bilayer} - E_{up} - E_{low}$ (where $E_{bilayer}$, $E_{up}$, and $E_{low}$ are the total energies of the bilayer and of the isolated upper and lower monolayers, respectively), is calculated to be −254.2 meV per formula unit (f.u.). The negative value indicates that the bilayer is energetically more favorable than the sum of the two isolated monolayers, confirming that the stacking is thermodynamically stable.

Table I. Structural parameters and key electronic properties of bilayer $VX_2$ (X = S, Se, and Te), including lattice constant $a = b$ (Å), interlayer distance $d$ (Å), binding energy $E_b$ (meV), bandgap $E_{gap}$ (meV), $P_z$ (pC/m) and VP (meV). $P_z$ and VP are listed separately for the AB and BA stackings, while the remaining parameters are identical for both stackings and are therefore given only once. The variation of $P_z$ with sliding distance for $VS_2$ and $VTe_2$ is shown in Fig. S3 of SM, revealing the trend $P_z(VS_2) < P_z(VSe_2) < P_z(VTe_2)$, while their valley-polarized band structures are shown in Fig. S4 of SM.

| System | $a = b$ | $d$ | $E_b$ | $E_{gap}$ | $P_z$ (AB) | $P_z$ (BA) | VP (AB) | VP (BA) |
|---|---|---|---|---|---|---|---|---|
| $VS_2$ | 3.18 | 3.11 | −219.2 | 93 | 0.57 | −0.57 | 54 | −54 |
| $VSe_2$ | 3.34 | 3.24 | −254.2 | 296 | 0.68 | −0.68 | 71 | −71 |
| $VTe_2$ | 3.58 | 3.38 | −312.5 | 85 | 1.00 | −1.00 | 95 | −95 |

Sliding is applied to the upper layer while the lower layer is fixed [see Fig. S1(b) of SM], with the displacement along the unit-cell diagonal, $\boldsymbol{x} = x(\boldsymbol{a} + \boldsymbol{b})$, where $\boldsymbol{a}$ and $\boldsymbol{b}$ are the lattice vectors of the hexagonal lattice and $x$ parametrizes the sliding distance. For $x$ = 0, 1/3, 1/2, and 2/3, the AA, AB, domain-wall (DW), and BA stackings are obtained, respectively. Interlayer AFM and interlayer FM orders are considered, with both layers intralayer FM. Along the entire sliding path, the stacking energy in the AFM order remains lower than that in the FM order, identifying AFM as the magnetic ground state, as shown in Fig. 2(a). This AFM preference arises from interlayer virtual hopping through Se-$p$ orbitals: antiparallel spins allow more hopping channels and lower the kinetic exchange energy, whereas parallel spins are partially blocked by the Pauli principle. The magnetic moments are mainly on V atoms (1.08 $\mu_B$) and fractionally on Se atoms (0.09 $\mu_B$). Corresponding atoms in the two layers carry equal but opposite moments (i.e. $M_{up} = -M_{low} = 1.26\ \mu_B$), giving zero net moment and indicating a fully compensated AFM state. The energy profile shows a double-well-like landscape, with two favorable minima corresponding to the AB and BA stackings. These are connected through an intermediate DW stacking, while the AA stacking lies at higher energy. The switching barrier between AB and BA is $E_{DW} - E_{AB(BA)} \approx 12.38$ meV/f.u.. This barrier is comparable to those previously reported for other 2D sliding ferroelectrics, with $YI_2$ (13.6 meV) [40], BN (9.0 meV) [52], $MoS_2$ (15.0 meV) [53], $VN_2(AlO)_2$ (13.1 meV) [54], and $VSi_2P_4$ (9.0 meV) [55], respectively, suggesting that these stackings are experimentally switchable.

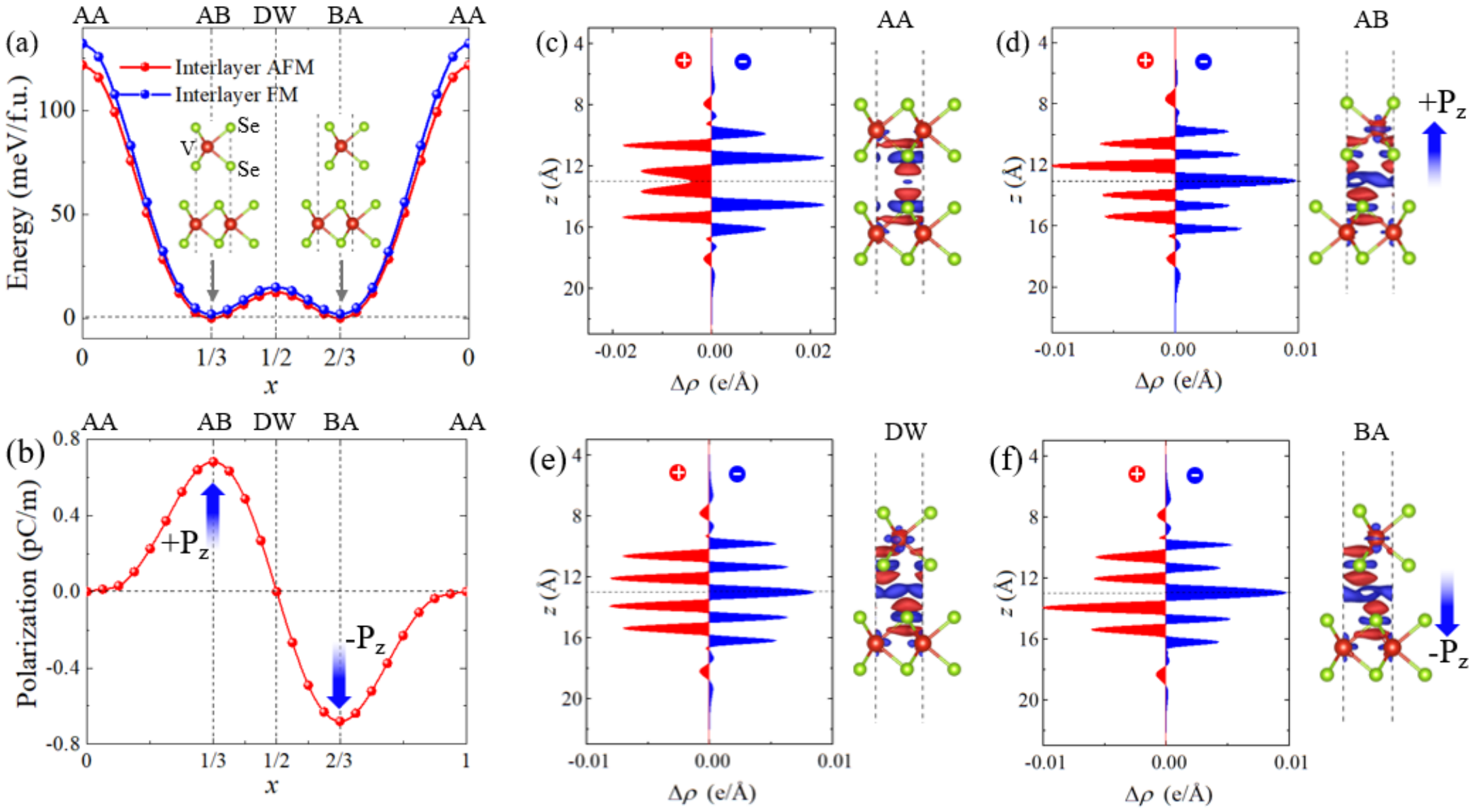

**Fig. 2**. Interlayer-sliding-induced ferroelectric polarization in bilayer $VSe_2$. (a) Dependence of stacking energy on sliding displacement $x$ along the diagonal of the unit cell, where interlayer antiferromagnetic (AFM) and ferromagnetic (FM) configurations show a double-well potential. Insets are atomic structures of the AB and BA stackings. (b) Sliding dependence of polarization magnitude, giving opposite signs in AB and BA. (c–f) Charge transfer (left panel) along $z$, obtained from the differential charge density (right panel), corresponding to (c) AA, (d) AB, (e) domain wall (DW), and (f) BA, respectively. Blue and red correspond to electron accumulation and depletion (i.e. holes), respectively.

Next, we discuss ferroelectric polarization from the symmetry constraints imposed by interlayer stacking. The AA-stacked bilayer belongs to the structural point group $D_{3h}$, which contains both the horizontal mirror operation $M_z$ and the threefold rotation $C_3$ about the out-of-plane axis. Under $M_z$, the polarization transforms as

$$M_z: (P_x, P_y, P_z) \rightarrow (P_x, P_y, -P_z). \quad (1)$$

or equivalently, $M_z P_z = -P_z$, while $P_x$ and $P_y$ remain unchanged. Because the AA structure is invariant under $M_z$, its out-of-plane polarization must satisfy $P_z = -P_z$, yielding $P_z = 0$. Meanwhile, the $C_3$ rotation transforms the in-plane polarization components as

$$C_3: (P_x, P_y) \rightarrow \left(-\frac{P_x}{2} - \frac{\sqrt{3}P_y}{2}, \frac{\sqrt{3}P_x}{2} - \frac{P_y}{2}\right). \quad (2)$$

Requiring the polarization to remain invariant under $C_3$ gives $P_x = P_y = 0$. Thus, the combined symmetry constraints of $D_{3h}$ forbid both in-plane and out-of-plane polarization in the AA stacking. In the AFM order, the combined magnetic symmetry $M_z T$, where $T$ denotes time reversal, also preserves the magnetic arrangement: $M_z$ exchanges the upper and lower layers, while $T$ reverses their magnetic moments. Since electric polarization is even under time reversal, $M_zT$ transforms $P_z$ into $-P_z$ and therefore also enforces $P_z = 0$. Upon sliding to AB or BA, the horizontal mirror symmetry $M_z$ is broken, whereas the threefold rotation $C_3$ is retained, reducing the structural point group from $D_{3h}$ to $C_{3v}$ (see Table II). The loss of $M_z$ removes the symmetry constraint that forbids $P_z$, while the retained $C_3$ rotation continues to require $P_x = P_y = 0$. Consequently, the AB and BA stackings allow a finite out-of-plane polarization but forbid in-plane polarization. Although AB and BA belong to the same point group, they are related by the horizontal mirror operation and thus correspond to oppositely polarized states, satisfying $P_z(\mathrm{AB}) = -P_z(\mathrm{BA})$. This symmetry evolution explains how interlayer sliding can generate switchable out-of-plane ferroelectricity while preserving the threefold rotational symmetry.

The calculated polarization profile is consistent with this symmetry analysis, as shown in Fig. 2(b). The out-of-plane polarization evolves continuously with the sliding displacement, vanishes at the AA stacking, and reaches approximately +0.68 pC/m and −0.68 pC/m for the AB and BA stackings, respectively. Although these values are smaller than those of conventional oxide ferroelectrics[56-58], they exceed the values reported for other van der Waals ferroelectrics and should therefore be experimentally detectable [59-62]. The opposite polarizations confirm that reversing the interlayer registry reverses the polarization direction. Combined with the double-well energy landscape, these results establish that the two polar configurations are energetically accessible states. The symmetry evolution and polarization profile therefore provide complementary evidence for sliding ferroelectricity.

A contrasting situation occurs when upper layer is rotated by $C_2$ while lower layer remains fixed [see Fig. S2(a) of SM]. For the considered rotated bilayer, the combined space-time inversion symmetry $PT$ is preserved along the interlayer-sliding pathway, where $P$ denotes spatial inversion. Because electric polarization is a polar vector, inversion reverses its direction, whereas time reversal leaves it unchanged. Therefore, $PT$ transforms the polarization as

$$PT: (P_x, P_y, P_z) \rightarrow (-P_x, -P_y, -P_z). \tag{3}$$

If the structure remains invariant under $PT$, its polarization must satisfy $(P_x, P_y, P_z) = (-P_x, -P_y, -P_z)$, requiring $P_x = P_y = P_z = 0$. Since interlayer sliding does not break $PT$ in this rotated bilayer, ferroelectric polarization remains symmetry-forbidden along the considered pathway, consistent with our first-principles calculations [see Fig. S2(b) of SM]. This comparison demonstrates that interlayer sliding alone does not guarantee ferroelectricity; the decisive factor is whether the sliding process breaks the symmetry that forbids the relevant polarization component.

Table II. Analysis of corresponding space group and point group from the high-symmetry relative stackings in bilayer $VX_2$ (X = S, Se and Te). Rotated stackings always preserve inversion symmetry and, when in the AFM order, possess the combined space-time inversion symmetry $PT$.

| Stacking | AA | AB | DW | BA |
|---|---|---|---|---|
| Direct | $P\bar{6}m2$ (No. 187): $D_{3h}$ | $P3m1$ (No. 156): $C_{3v}$ | $Abm2$ (No. 39): $C_{2v}$ | $P3m1$ (No. 156): $C_{3v}$ |
| $C_2$ rotating | $P\bar{3}m1$ (No. 164): $D_{3d}$ | $P\bar{3}m1$ (No. 164): $D_{3d}$ | $C2/m$ (No. 12): $C_{2h}$ | $P\bar{3}m1$ (No. 164): $D_{3d}$ |

To elucidate the microscopic origin of the polarization, we further analyze the differential charge density, expressed by

$$\Delta\rho(\boldsymbol{r}) = \rho_{\mathrm{bilayer}}(\boldsymbol{r}) - \rho_{\mathrm{up}}(\boldsymbol{r}) - \rho_{\mathrm{low}}(\boldsymbol{r}), \tag{4}$$

where $\rho_{\mathrm{bilayer}}$ denotes the total charge density of the bilayer, while $\rho_{\mathrm{up}}$ and $\rho_{\mathrm{low}}$ refer to the charge densities of the upper and lower monolayers calculated separately in the same cell and at the same atomic coordinates as the bilayer. The corresponding charge transfer along the out-of-plane direction is calculated by

$$\Delta\rho(z) = \iint \Delta\rho(\boldsymbol{r})\, dx\, dy. \tag{5}$$

This expression describes the spatial redistribution of electrons in the interlayer region upon bonding; $\Delta\rho(z) > 0$ corresponds to electron accumulation, while $\Delta\rho(z) < 0$ denotes electron depletion (i.e. holes). The calculated results for the AA, AB, DW, and BA stackings are shown in Figs. 2(c)–2(f). The AA stacking exhibits a nearly symmetric charge distribution with respect to the horizontal mirror plane, consistent with its vanishing net polarization. Although local charge redistribution occurs, its contributions to the total dipole cancel by symmetry. A similar cancellation is observed for the DW stacking, where the calculated polarization also vanishes. In contrast, the AB stacking exhibits an asymmetric distribution of charge accumulation and depletion along the out-of-plane direction. The change in interlayer registry modifies the local atomic environment and breaks the equivalence between the two layers, producing a nonzero net dipole moment. Upon sliding from AB to BA, the charge redistribution reverses its out-of-plane asymmetry, resulting in an opposite polarization direction. These results show that interlayer charge redistribution alone is insufficient to generate a macroscopic dipole; the spatial asymmetry of the redistributed charge is essential. This interpretation is consistent with the symmetry reduction from $D_{3h}$ to $C_{3v}$: the loss of the layer-exchanging horizontal mirror symmetry permits an asymmetric charge distribution and a finite out-of-plane polarization.

### C. Layer-valley locking and ferroelectric switching of VP

Having established the symmetry origin of sliding ferroelectricity, we next investigate how it controls the valley splitting. Figures 3(a)–3(d) show the layer-projected band structures, including SOC, for the AA, DW, AB, and BA stackings, respectively. The valence-band-edge states at the two inequivalent valleys, $K_1$ and $K_2$, exhibit pronounced layer-dependent character. However, the valley degeneracy depends on the stackings: the two valleys remain degenerate in the nonpolar AA and DW, whereas finite valley splittings emerge in the polar AB and BA. Defining the valence-band (VB)-valley splitting as

$$\Delta E_{\mathrm{VP}} = E_{\mathrm{VB}}(K_1) - E_{\mathrm{VB}}(K_2), \tag{6}$$

we obtain a VP of $\Delta E_{\mathrm{VP}} \approx +71$ meV for AB and $-71$ meV for BA. This strong VP far exceeds the room-temperature thermal energy $K_{\mathrm{B}}T = 25.9$ meV [25], effectively suppressing thermal fluctuations and thereby enabling stably experimental observation of VP above room temperature. The reversal of VP demonstrates that interlayer sliding reverses the valley order. Importantly, this switching occurs without reversing the AFM order, indicating that VP can be controlled through ferroelectric switching.

To elucidate this mechanism, we catch the layer polarization of the valley states. For a Bloch state $\psi_{n\boldsymbol{k}}(\boldsymbol{r})$, the layer polarization $P_l(n, \boldsymbol{k})$ is defined by the probability weight in the upper layer minus that in the lower layer, namely,

$$P_l(n, \boldsymbol{k}) = W_{\mathrm{up}}(n, \boldsymbol{k}) - W_{low}(n, \boldsymbol{k}), \tag{7}$$

where $W_{\mathrm{up}}$ of upper layer and $W_{\mathrm{low}}$ of lower layer are obtained by integrating the electronic density $|\psi_{n\boldsymbol{k}}(\boldsymbol{r})|^2$ over the respective layer regions. Thus, $P_l > 0$ indicates that a state has more weight in the upper layer, whereas $P_l < 0$ indicates predominant localization in the lower layer. The magnitude $|P_l|$ measures the degree of layer polarization: $|P_l| = 1$ corresponds to complete localization in one layer, while $P_l = 0$ indicates even electronic distribution in the two layers. As shown in Figs. 3(e) and 3(f), the spatial distributions of the valence-band-edge states reveal that the $K_1$ electronic state is predominantly localized in the upper layer, whereas the $K_2$ electronic state is mainly localized in the lower layer. Accordingly, $P_l(K_1) > 0$ and $P_l(K_2) < 0$ establish valley-contrasting layer polarization. This behavior constitutes layer-valley locking, in which the valley degree of freedom is correlated with the layer distribution of the corresponding Bloch states. Notably, the layer character of each valley remains largely preserved upon switching between AB and BA. Therefore, the reversal of VP does not require the two valley states to exchange their layer localization.

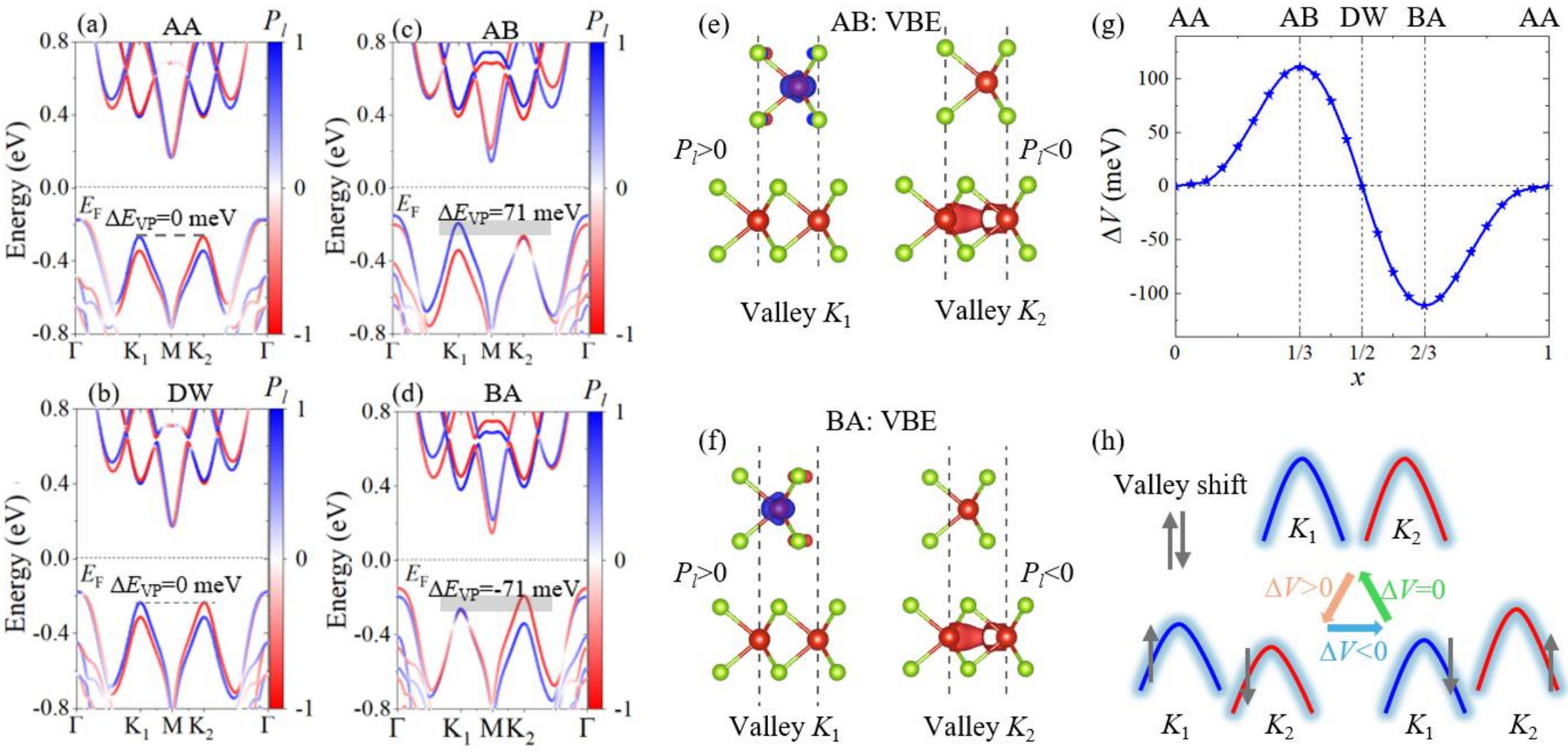

**Fig. 3**. (a–d) Layer-projected band structures with spin-orbital coupling (SOC), exhibiting pronounced valley-contrasting layer polarization ($P_l$). Both (a) AA and (b) DW stackings are valley-degenerate, whereas both (c) AB and (d) BA stackings possess opposite VPs. (e, f) Bloch electronic distribution of valence band edge (VBE)-valleys $K_{1,2}$ in AB and BA stackings. (g) Dependence of layer potential $\Delta V$ on sliding displacement. (h) Illustration of the valley shift caused by the coupling of $\Delta V$ to layer polarization.

Instead, their relative energies are reversed by the sliding-induced electrostatic potential. To understand this, we define the layer potential as the effective electrostatic potential difference between the upper and lower layers, $\Delta V = V_{\text{up}} - V_{\text{low}}$. The coupling between this layer potential and the layer polarization shifts the Bloch states. Specifically, for a valley index $\tau$, the energy shift is approximately $\delta E_\tau \approx \frac{\Delta V}{2} P_l(\tau)$, so states with opposite layer polarizations experience opposite energy shifts. The resulting valley splitting is approximately

$$\Delta E_{\text{VP}} \approx \frac{\Delta V}{2}[P_l(K_1) - P_l(K_2)]. \tag{8}$$

This relation establishes the connection between the real-space electrostatic asymmetry and the momentum-space valley splitting. The layer potential determines the magnitude and sign of the layer-dependent energy bias, while the difference in layer polarization between $K_1$ and $K_2$ determines how differently the two valleys respond to that bias. Consequently, VP is governed by the interplay between the layer potential and the valley-contrasting layer polarization, rather than by either quantity alone. As shown in Fig. 3(g), $\Delta V$ vanishes in the nonpolar AA and DW stackings and develops opposite signs in the AB and BA stackings, with a magnitude of approximately 111 meV (see Fig. S5 of SM for details). This evolution follows the ferroelectric polarization reversal. Crucially, the layer-polarization contrast retains its sign. Reversing the ferroelectric polarization therefore reverses the sign of $\Delta V$ while largely preserving $P_l(K_1) - P_l(K_2)$. From Equation 8, this reversal changes the relative energy shifts of the two valleys and consequently reverses $\Delta E_{\mathrm{VP}}$, consistent with the calculated splittings of approximately +71 meV for AB and −71 meV for BA. For the nonpolar stackings, the vanishing layer potential is consistent with the degeneracy of the two valleys. Figure 3(h) captures the essential switching mechanism: sliding ferroelectricity reverses the layer potential, and because the valley-contrasting layer polarization is preserved, this reversal is converted into a switch of the valley order. By contrast, in conventional ferrovalley materials, VP switching is achieved by magnetic-moment flipping under magnetic-valley locking [63, 64]. It therefore follows that, in the present system, sliding ferroelectric enables the reversal of VP without reversing magnetic order. The calculated VP and layer potential further allow us to estimate the effective magnitude of the layer polarization. Using $|\Delta E_{\mathrm{VP}}| \approx 71$ meV and $|\Delta V| \approx 111$ meV gives $P_l(K_1) - P_l(K_2) \approx \frac{2|\Delta E_{\mathrm{VP}}|}{|\Delta V|} \approx 1.28$. If the two valleys have approximately equal and opposite layer polarizations, this estimate yields $P_l(K_1) \approx +0.64$ and $P_l(K_2) \approx -0.64$. The estimate therefore indicates that the layer polarization at the valence-band edges is substantial but incomplete.

## D. Ferroelectric switching of the layer-locked AVHE

Having established that sliding ferroelectricity reverses the valley order through layer-valley locking, we next investigate its consequences for anomalous transverse transport. Figures 4(a) and 4(b) show the Wannier-interpolated band structures for the AB and BA stackings, respectively, together with the corresponding DFT results. The excellent agreement between the Wannier-fitted and DFT bands demonstrates that the Wannier tight-binding models accurately reproduce the band structures of both ferroelectric states. This agreement supports the use of the Wannier representation for subsequent transport calculations. We apply the Kubo formula to express the out-of-plane BC of a summation of all occupied states [65, 66]:

$$\Omega_z(\boldsymbol{k}) = -\sum_n \sum_{m\neq n} f_n \frac{2\hbar^2 \mathrm{Im}\langle\psi_{n\boldsymbol{k}}|\hat{v}_x|\psi_{m\boldsymbol{k}}\rangle\langle\psi_{m\boldsymbol{k}}|\hat{v}_y|\psi_{n\boldsymbol{k}}\rangle}{(E_{m\boldsymbol{k}}-E_{n\boldsymbol{k}})^2}, \tag{9}$$

where $f_i$ is the Fermi-Dirac distribution function, $|\psi_{i\boldsymbol{k}}\rangle$ and $E_{i\boldsymbol{k}}$ denote the eigenstate and eigenenergy of Bloch electrons, as well as $\hat{v}_i$ is the velocity operators obtained from the partial derivatives of the Wannier-interpolated Hamiltonian. The BC over the 2D Brillouin zone is calculated for the AB stacking, predominantly concentrated around the inequivalent valleys $K_1$ and $K_2$, as shown in Fig. 4(c). The two valleys exhibit opposite signs and nearly equal magnitudes, namely, $\Omega_z(K_1) \approx -\Omega_z(K_2)$. This implies opposite anomalous transverse velocities under an in-plane electric field. Specifically, the anomalous velocity is given by [2]

$$\boldsymbol{v}_\perp(\boldsymbol{k}) = \dot{\boldsymbol{k}} \times \Omega_z(\boldsymbol{k}), \tag{10}$$

where $\hbar\dot{\boldsymbol{k}} = -e\boldsymbol{E}$ for electrons, with $\boldsymbol{E}$ being the in-plane electric field and $e$ the elementary charge. Thus, the valley-contrasting BC provides the microscopic basis for the anomalous transverse response. For the BA stacking, the valley-contrasting patterns persist, as shown in Fig. 4(d). Thus, switching between AB and BA does not induce a reconstruction of BC, suggesting that the ferroelectric control of the Hall response originates primarily from changes in valley energetics and occupation.

To quantify the resulting transport response, we calculate the intrinsic AHC as a function of the chemical potential, $\mu$. It is calculated by integrating the BC over the occupied electronic states [67]:

$$\sigma_{xy}(\mu) = \frac{e^2}{\hbar}\sum_n \int_{\mathrm{BZ}} \frac{d^2k}{(2\pi)^2} f_n(\mu)\Omega_n^z(\boldsymbol{k}). \tag{11}$$

This expression makes explicit that the AHC depends on both the BC distribution and the occupation of the corresponding electronic states. As shown in Figs. 4(e) and 4(f), the evolution of AHC with chemical potential reveals distinct transport regimes. When the chemical potential lies within the band gap, the AHC is nearly zero for both stackings. As it moves downward into the valence-band region, the Γ-centered bands give rise to a small initial increase. With further lowering of the chemical potential, this contribution gradually decreases and eventually returns to nearly zero before the chemical potential reaches the VP window. This nonmonotonic behavior indicates that the Γ-centered states do not provide a sustained net transverse response. A pronounced AHC emerges only when the chemical potential enters the VP window, where the valley-resolved valence-band edges become selectively depopulated. Despite their similar BC distributions, the two stackings exhibit opposite AHC within the shaded VP window [i.e., $\sigma_{xy}^{\mathrm{AB}}(\mu) = -\sigma_{xy}^{\mathrm{BA}}(\mu)$]: the AB stacking shows a negative conductivity, whereas the BA stacking exhibits a positive one.

To clarify the origin of this response, the total conductivity can be separated into contributions from the two valley regions and the remainder of the Brillouin zone:

$$\sigma_{xy}(\mu) = \sigma_{xy}^{K_1}(\mu) + \sigma_{xy}^{K_2}(\mu) + \sigma_{xy}^{\mathrm{rest}}(\mu). \tag{12}$$

Here, $\sigma_{xy}^{K_1}$ and $\sigma_{xy}^{K_2}$ denote the BC contributions integrated over momentum-space regions surrounding $K_1$ and $K_2$, respectively, while $\sigma_{xy}^{\mathrm{rest}}$ represents the contribution from the remaining states. Although the states outside the two valley regions provide finite contributions, their positive and negative Berry-curvature contributions largely cancel within the VP window, leaving a relatively small residual background. Consequently, the net Hall response is dominated by $K_1$ and $K_2$. Since these two valleys possess opposite BCs, their occupation-weighted contributions tend to compensate when their magnitudes are comparable. Within the VP window, however, the valley-selective hole occupation prevents complete cancellation, resulting in a finite AHC. The opposite Hall responses of AB and BA can be understood from their reversed valley order. For the AB stacking, the $K_1$ valley lies higher in energy than $K_2$ at the valence-band edge. Consequently, holes preferentially emerge around $K_1$ as the chemical potential enters the VP window. Since $K_1$ carries positive BC, the removal of occupied electronic states around this valley produces a negative change in the AHC. In contrast, the BA stacking exhibits the opposite valley order, with $K_2$ lying higher in energy than $K_1$. Holes therefore preferentially emerge around $K_2$, whose BC is negative, resulting in a positive change in the AHC. Thus, switching between AB and BA reverses the relative valley contributions and hence the sign of the net Hall response, in contrast to the conventional ferrovalley strategy, where reversing the AHC requires magnetization reversal and concomitant reconstruction of the BC distribution [27].

Finally, the layer-resolved electronic states establish the connection between this reversible Hall response and the ferroelectric polarization. Because $K_1$ and $K_2$ are predominantly localized in different layers, switching the hole valley from $K_1$ in AB to $K_2$ in BA also switches the layer associated with the Hall-active states. This constitutes a layer-locked AVHE, in which ferroelectric switching reverses the sign and layer character of the transverse response, as shown in Fig. 1(b). Unlike the classical, spin, valley, and anomalous Hall effects [68-70], each of which is associated with only one degree of freedom, this AFM bilayer enables a single transport process to engage charge, spin, valley, and layer degrees of freedom at the same time. Such intrinsic locking among layer, valley, and spin degrees of freedom renders the system highly attractive for electronics, layertronics, spintronics, valleytronics, and their interdisciplinary applications.

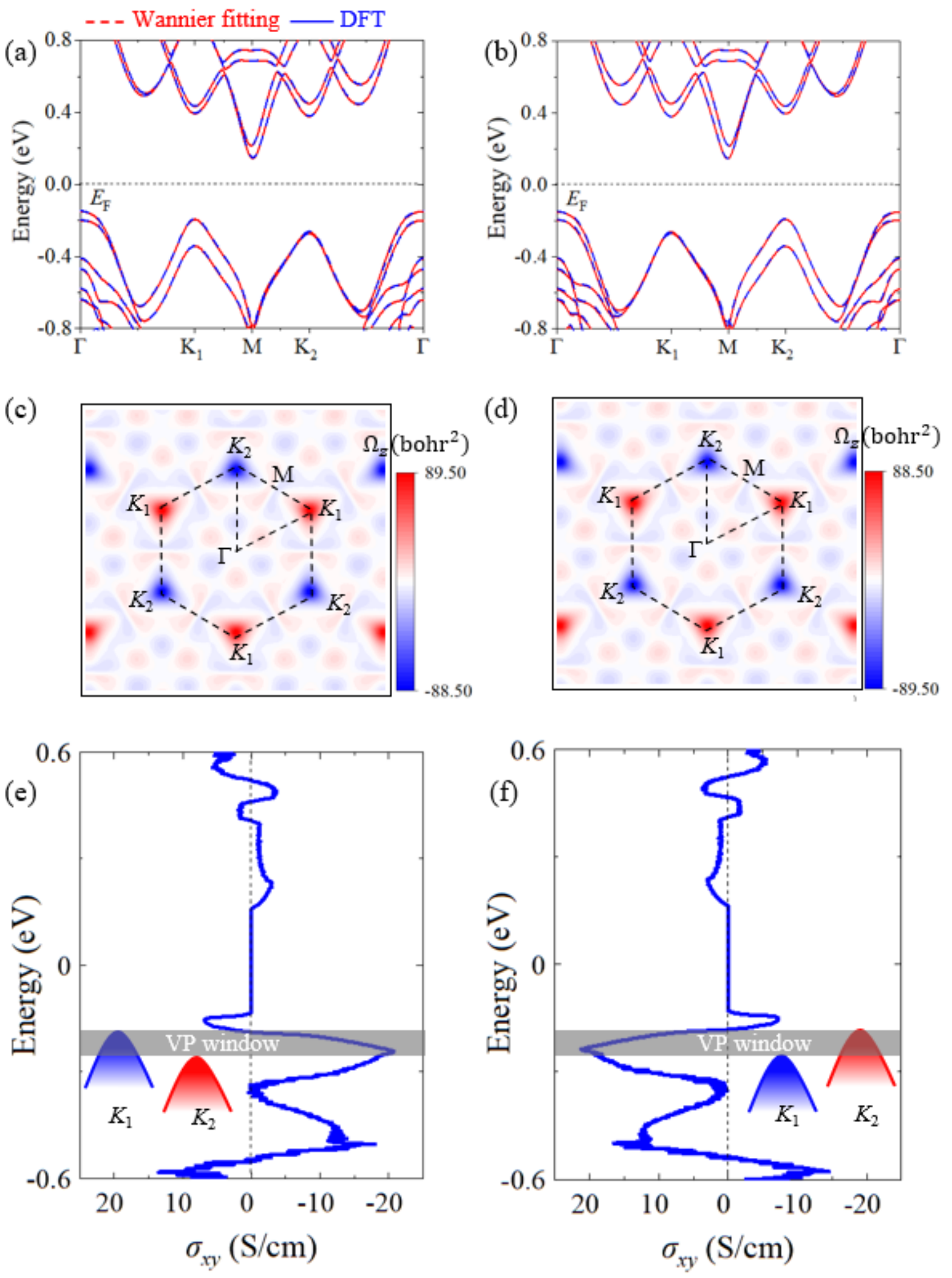

**Fig. 4**. Anomalous transport calculations for both (a, c, e) AB and (b, d, f) BA configurations. (a, b) DFT band dispersions fitted by a tight-binding model constructed from Wannier functions. Red dashed lines and blue solid lines denote the Wannier fitting results and DFT scheme, respectively. (c, d) Contour maps of BC over the entire 2D Brillouin zone, which is primarily localized around the valleys $K_{1,2}$. (e, f) Variation of anomalous Hall conductivity with the Fermi level. The shaded region marks the VP window that gives rise to AVHE.

### E. Robustness of ferroelectric-valley-switching mechanism against electronic correlation

To assess the robustness of the proposed ferroelectric-valley-switching mechanism against electronic correlation, we investigate the dependence of magnetic stability, bandgap, ferroelectric polarization, and VP on the effective Hubbard parameter $U_{\mathrm{eff}}$. As shown in Fig. 5(a), the magnetic energy difference $E_{\mathrm{AFM}} - E_{\mathrm{FM}}$ is positive at $U_{\mathrm{eff}} = 0$, indicating that the FM order is energetically favored in the absence of the Hubbard correction. With increasing $U_{\mathrm{eff}}$, it becomes negative, and the AFM order becomes energetically favorable. This crossover demonstrates that electronic correlation plays a crucial role in stabilizing the A-type AFM ground state considered in this work. The bandgap also depends nonmonotonically on $U_{\mathrm{eff}}$, increasing from approximately 0.09 eV at $U_{\mathrm{eff}} = 0$ to a maximum of approximately 0.46 eV near $U_{\mathrm{eff}} = 2$ eV, before decreasing at larger $U_{\mathrm{eff}}$, as shown in Fig. 5(b). Nevertheless, a finite bandgap persists throughout the investigated parameter range, indicating that the system remains semiconducting within the considered calculations.

We next examine whether the sliding-induced ferroelectric polarization is sensitive to the strength of electronic correlation. The calculated out-of-plane polarizations of the AB and BA stackings retain opposite signs throughout the investigated $U_{\mathrm{eff}}$ range, as shown in Fig. 5(c). For AB stacking, the polarization increases from approximately 0.3 pC/m at $U_{\mathrm{eff}} = 0$ and approaches 0.7 pC/m at larger $U_{\mathrm{eff}}$. The BA stacking exhibits the opposite polarization, with a comparable magnitude. Although the absolute polarization changes with $U_{\mathrm{eff}}$, its sign contrast between AB and BA remains robust with

$$P_z^{\mathrm{AB}}(U_{\mathrm{eff}}) = -P_z^{\mathrm{BA}}(U_{\mathrm{eff}}). \tag{13}$$

This behavior indicates that the reversal of ferroelectric polarization is primarily determined by the stacking associated structural asymmetry, rather than being restricted to a particular value of the Hubbard parameter. VP exhibits a similarly robust response to electronic correlation (see Fig. S6 of SM). The calculated splitting still satisfies

$$\Delta E_{\mathrm{VP}}^{\mathrm{AB}}(U_{\mathrm{eff}}) = -\Delta E_{\mathrm{VP}}^{\mathrm{BA}}(U_{\mathrm{eff}}), \tag{14}$$

remaining positive for AB stacking and negative for BA stacking throughout the investigated $U_{\mathrm{eff}}$ range, as shown in Fig. 5(d). Its magnitude increases from approximately 45 meV at $U_{\mathrm{eff}} = 0$ to around 75 meV over much of the finite-$U_{\mathrm{eff}}$ range, with some variation at larger $U_{\mathrm{eff}}$. Thus, the valley order remains opposite for the two stackings, while the magnitude of the splitting is moderately changed by electronic correlation. Particularly, once the AFM order becomes energetically favorable, VP remains substantial and preserves the sign reversal associated with switching between AB and BA. The correlated evolution of ferroelectric polarization and VP provides further support for the mechanism that interlayer sliding determines the sign of the ferroelectric polarization and the associated layer potential, while layer-valley locking converts this potential asymmetry into a valley-dependent energy shift. Consequently, reversing the stacking reverses the valley order even as $U_{\mathrm{eff}}$ modifies the values of the ferroelectric polarization and VP. The persistence of opposite polarization and VP signs therefore indicates that the coupling between sliding ferroelectricity and valley splitting is not confined to a narrowly selected correlation strength.

Taken together, these results show that electronic correlation influences the magnetic ground state and quantitatively renormalizes the bandgap, ferroelectric polarization, and VP, but does not eliminate the contrasting ferroelectric and VP. The results support the robustness of the proposed sliding-ferroelectric control of VP in the regime where the A-type AFM configuration is energetically stable. Combined with the associated valley transport, this robustness provides a basis for the ferroelectric switching of the layer-locked anomalous valley Hall response.

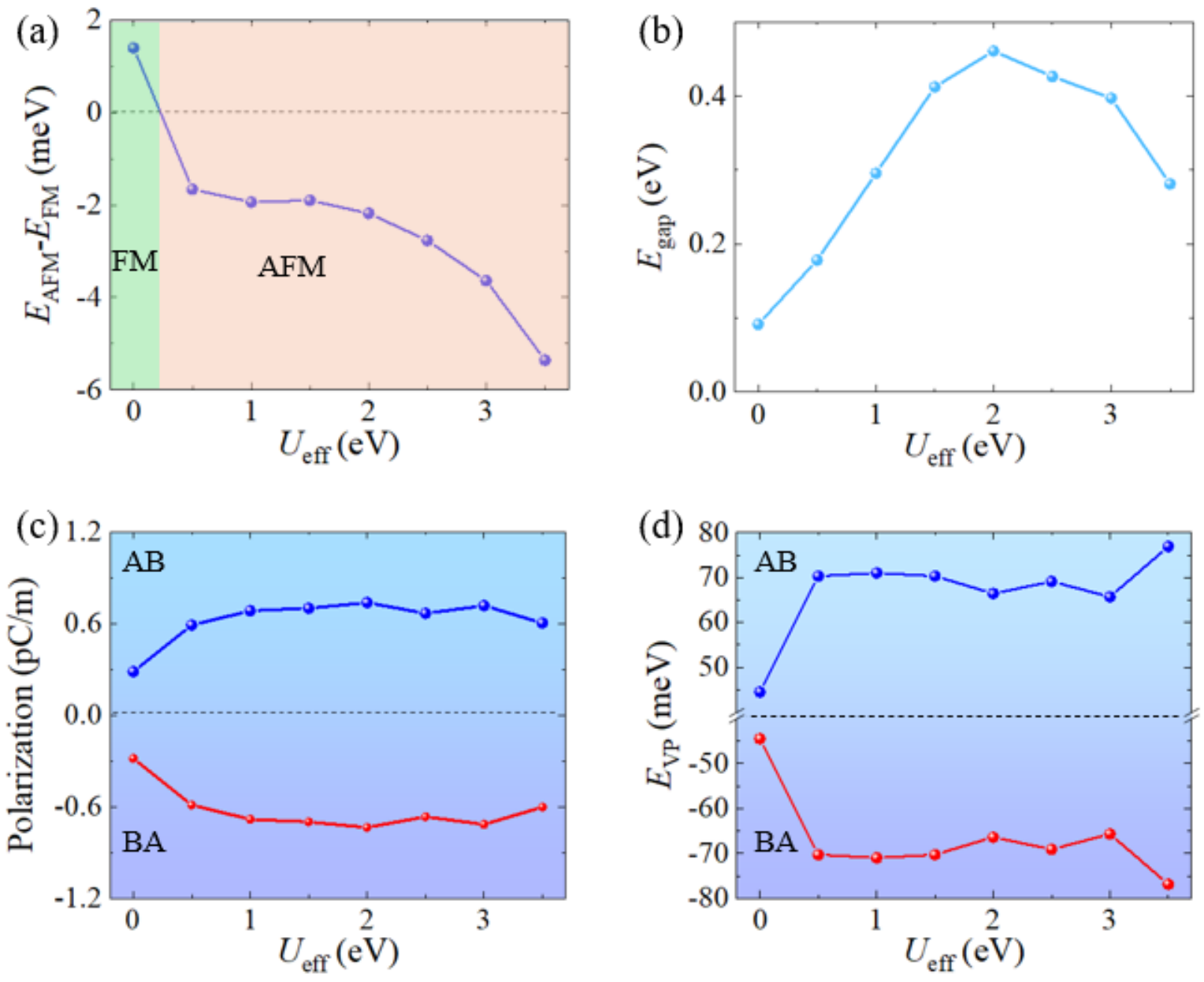

**Fig. 5**. Influence of the effective electronic correlation strength ($U_{eff}$) on key electronic parameters. (a) Energy difference between the interlayer AFM and FM configurations versus $U_{eff}$, exhibiting a magnetic phase transition. (b) Bandgap dependence on $U_{eff}$. Variation of (c) ferroelectric polarization and (d) VP with $U_{eff}$ for both AB and BA stackings.

## IV. CONCLUSION

In summary, first-principles calculations and symmetry analysis reveal a sliding-ferroelectric mechanism for switching VP and the layer-locked AVHE in bilayer $VSe_2$ with A-type AFM order. Interlayer sliding lowers the point-group symmetry from $D_{3h}$ to $C_{3v}$, breaking the horizontal mirror symmetry but preserving $C_3$ rotational symmetry. This allows a pronounced out-of-plane ferroelectric polarization arising from asymmetric interlayer charge redistribution, while in-plane polarization remains prohibited. Building on this symmetry-enabled ferroelectricity, we demonstrate that the sliding-induced layer potential couples to the valley-contrasting layer polarization of the valence-band-edge states, reversing the valley order and enabling substantial, switchable VP without reversing the AFM order. Furthermore, since the two valleys possess opposite BCs, this reversal of valley order changes their relative occupation and switches the sign of the AVHE, while the BC distribution remains largely preserved. Both VP and the layer-locked AVHE can therefore be reversibly controlled by ferroelectric switching, offering the prospect of nonvolatile valley encoding and Hall readout. The persistence of opposite ferroelectric polarization and VP over effective Hubbard parameters further supports the robustness of this mechanism. These findings establish interlayer sliding as a promising strategy for integrating ferroelectric switching, valley manipulation, and transverse transport control, providing new opportunities for nonvolatile valleytronics in compensated AFM materials.

## ACKNOWLEDGMENTS

This work is supported by the National Natural Science Foundation of China (Grant No.12674079). We also acknowledge the Fermi cluster at Lanzhou University for providing computational resources.